\documentclass[hidelinks,conference]{IEEEtran}
\IEEEoverridecommandlockouts
\usepackage{cite}
\usepackage{amsmath,amssymb,amsfonts}
\usepackage{algorithmic}
\usepackage{graphicx}
\usepackage{textcomp}
\usepackage{xcolor}
\usepackage{subfigure}
\usepackage{hyperref}
\def\BibTeX{{\rm B\kern-.05em{\sc i\kern-.025em b}\kern-.08em
    T\kern-.1667em\lower.7ex\hbox{E}\kern-.125emX}}

\makeatletter
\newcommand{\linebreakand}{%
  \end{@IEEEauthorhalign}
  \hfill\mbox{}\par
  \mbox{}\hfill\begin{@IEEEauthorhalign}
}
\makeatother

\begin{document}

\title{Log Parsing Evaluation in the Era of Modern Software Systems}

\author{
\IEEEauthorblockN{Stefan Petrescu}
\IEEEauthorblockA{\textit{Dependable and Scalable Software Systems}\\
\textit{Leibniz University Hannover}\\
\href{mailto:petrescu@vss.uni-hannover.de}{petrescu@vss.uni-hannover.de}}\\   %
\IEEEauthorblockN{Alexandru Uta}
\IEEEauthorblockA{\textit{DFINITY} \\
\textit{Zurich, Switzerland}\\
\href{mailto:alexandru.uta@dfinity.org}{alexandru.uta@dfinity.org}}
\and
\IEEEauthorblockN{Floris den Hengst}
\IEEEauthorblockA{\textit{ING Analytics}\\
\textit{Amsterdam, the Netherlands}\\
\href{https://orcid.org/0000-0002-2092-9904}{0000-0002-2092-9904}}\\  %
\IEEEauthorblockN{Jan S. Rellermeyer}
\IEEEauthorblockA{\textit{Dependable and Scalable Software Systems}\\
\textit{Leibniz University Hannover}\\
\href{mailto:rellermeyer@vss.uni-hannover.de}{rellermeyer@vss.uni-hannover.de}}
}

\maketitle
\IEEEpeerreviewmaketitle

\begin{abstract}
Due to the complexity and size of modern software systems, the amount of logs generated is tremendous. Hence, it is infeasible to manually investigate these data in a reasonable time, thereby requiring automating log analysis to derive insights about the functioning of the systems. Motivated by an industry use-case, we zoom-in on one integral part of automated log analysis, \emph{log parsing}, which is the prerequisite to deriving any insights from logs. Our investigation reveals problematic aspects within the log parsing field, particularly its inefficiency in handling heterogeneous real-world logs. We show this by assessing the 14 most-recognized log parsing approaches in the literature using (i) nine publicly available datasets, (ii) one dataset comprised of combined publicly available data, and (iii) one dataset generated within the infrastructure of a large bank. Subsequently, toward improving log parsing robustness in real-world production scenarios, we propose a tool, \textsc{Logchimera}, that enables estimating log parsing performance in industry contexts through generating synthetic log data that resemble industry logs. Our contributions serve as a foundation to consolidate past research efforts, facilitate future research advancements, and establish a strong link between research and industry log parsing.
\end{abstract}


\section{Introduction}
\noindent
Logs record system runtime information that is crucial for assessing, predicting, and improving the reliability, safety, and security of software systems~\cite{nandi2016anomaly}. For example, logs are necessary for system monitoring~\cite{candido2021log}, alerting of errors and anomalies~\cite{8804456}, incident mitigation~\cite{nagaraj2012structured,oliner2012advances}, root cause analysis~\cite{5609556} and auditing~\cite{5254059}. These, however, are all challenging tasks that are expected to become more challenging in the future due to the increasing pervasiveness, complexity and size of software systems.

While log data is a key resource for important reliability engineering tasks, extracting information from log data is challenging: successfully implementing processes that automate (at least partially) log analysis can be difficult and resource-intensive, given the ever increasing systems' complexity and log data volumes generated (e.g., TBs of log data daily~\cite{https://doi.org/10.48550/arxiv.1910.00409}). As a result, it is not uncommon for logs to remain mostly unused~\cite{9970498}. In practice, the harsh but true reality is for logs to only be looked at in critical situations, where engineers still analyze logs manually~\cite{10.1145/3460345,8812113}. This classical `needle in a haystack under high pressure' scenario illustrates the need for automated approaches to make log analysis more applicable in modern production scenarios.

In this work, we zoom in on one integral part of automated log analysis known as \emph{log parsing}\footnote{`Log parsing' is sometimes used interchangeably with `log abstraction' or `event template extraction'.} (LP), a well-contained and first step in the log analysis process and therefore an excellent target for improving this process in practice. Specifically, motivated by an industry use-case of a large financial institute, we investigate the performance and applicability of log parsing solutions in literature and in the context of modern software systems in practice; we discuss our findings and propose solutions for solidifying the field and enhancing the way how log parsers can be evaluated in more realistic settings.

In our study, we discover some aspects of log parsing in literature that may limit its adoption and may surprise from a current and practical perspective. We list these aspects and propose improvements for these gaps to strengthen the fields' evaluation methodology, toward increasing adoption in modern production environments. We continue by listing these aspects and mention how we aim to address them in this work.

We first focus on the aspect of evaluation in literature and compare it to functional requirements of log parsing in industry. In literature, across many studies, an evaluation metric called \emph{parsing accuracy}~\cite{10.1145/3460345} is used, adopted as the standard for evaluation~\cite{8804456,DBLP:journals/corr/abs-2110-12489,ELMASRI2020106276,zhangsystem}. However, at a closer look, even though a valid metric, it provides estimates for something only marketed as log parsing. Whereas the name `parsing accuracy' implies that this metric measures how well a text string is processed into separate components, the de-facto definition of this metric in literature is the segmentation of text strings into different clusters. This, although relevant in scenarios where \emph{log clustering} is the goal, may actually provide misleading comparisons in literature for \emph{log parsing} and a poor translation of results in research and expectation in industry, leading to e.g., an accuracy of 73\% on average in literature~\cite{8804456} where only 20\% of target fields are identified in practice (Section~\ref{sec:accuracy}). This subtle difference has been also been noticed recently~\cite{Liu_2022,DBLP:journals/corr/abs-2003-07905}, and in our experiments we further showcase its relevance and the effects of accounting for it, by conducting an analysis of log parsing performance on (i) publicly available data, (ii) data that resembles industry logs, and on (iii) actual real-world industry logs. %

The second aspect of focus in this work is the representativeness of the evaluation in literature for actual software systems in industry today. In particular, we argue that evaluation data in literature stems from large-scale applications standardized on single technologies~\cite{DIFRANCESCO201977} (e.g., monolithic application) whereas modern software systems in industry today are comprised of flexible and easy-to-evolve architectures~\cite{lewis2014microservices} (e.g., microservices) which have a strong tendency to produce more heterogeneous log data as a result~\cite{Liu_2022,candido2021log}. 
We argue that log parsing evaluation data no longer fully reflect what is found in the context of contemporary real-world systems, e.g., some of the current evaluation datasets in the field are even comprised of 15 year-old logs~\cite{He2020LoghubAL}. Ideally, experiments should be conducted on production data but we acknowledge that organizations are reluctant to share such data with researchers, for instance, due to privacy, confidentiality, and security concerns.

In this work, we address the aforementioned issues by presenting a novel tool, \textsc{Logchimera}, that generates data from low-complexity public data sets like the ones used in the state-of-the-art literature and creates new datasets that resemble industry logs and thereby enable realistic comparisons of a variety of log parsing methods. We propose two techniques, mixing and fuzzing, that jointly allow us to approximate the complexity of real-world production logs. 

Our paper makes the following contributions: \textbf{(1)} An evaluation of log parsing considering metrics that aim to strengthen its connection and applicability to industry, on various log data, including real-world industry data collected from the infrastructure of a large financial institute; \textbf{(2)} An analysis of the characteristics of industry logs compared to publicly available data; \textbf{(3)} Two publicly available datasets for benchmarking: one for running log parsing experiments and one that enables generating synthetic data that contains similar characteristics to industry logs; \textbf{(4)} A tool, \textsc{Logchimera}, that enables creating data that resembles industry logs' heterogeneity, and testing on custom data with varied levels of heterogeneity, enabling obtaining estimates for log parsing's performance in industry contexts. Our contributions lay the foundation for uniting past research efforts, enabling future research efforts to compound, and creating a solid connection between research and industry log parsing.

All code and data used for the implementation as well as the experiments are available as open source under \texttt{\url{https://github.com/spetrescu/logchimera}}.

\section{Background and Related Work}
\noindent
Automated log analysis involves a sequence of steps that collectively aim at translating log data into actionable insights. This reduces the complexity of the overall process by tackling a number of intermediate (easier) problems that can lead to distilling critical information from the logs. The first step usually is to abstract away from runtime logs, as subsequent steps expect data to have a particular structure. To do that, most techniques require a basic syntax-derived exploration and interpretation of the logs~\cite{7266837}, known in the literature as \emph{log parsing}~\cite{8804456,7579781,ELMASRI2020106276,10.1145/3465481.3470083,9338336}. 

Log parsing transforms runtime logs by attempting to discover (1) the underlying log templates corresponding to the static part of the logging statements in the software, (2) their respective parameters corresponding to the dynamic part of the logging statements, and (3) by appending log meta information. Log meta-information, such as \texttt{PID}, \texttt{Date}, \texttt{Timestamp}, \texttt{Level}, \texttt{Component}, etc. is usually added by a logging framework~\cite{tao2021logstamp} and thus relatively easy to obtain~\cite{10.1145/3460345}. Consequently, the main challenge of log parsing is discovering log templates and parameters. 

\begin{figure}[hbt!]
	\centering
	\includegraphics[width=0.9\linewidth]{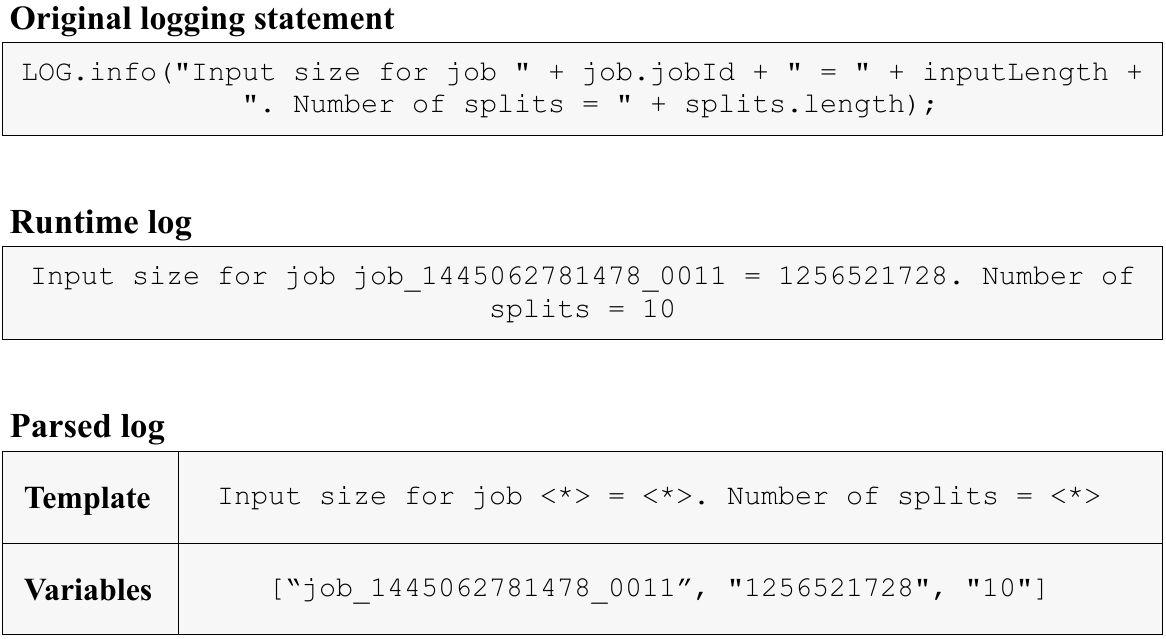}
	\caption{Example of log parsing process. The discovered constant parts represent the log template, whereas variables are replaced by generic tokens: `\texttt{<*>}'.}
	\label{fig:log_parsing_example_transformation}
\end{figure}

\begin{figure*}
	\centering
	\subfigure{\includegraphics[width=0.46\linewidth]{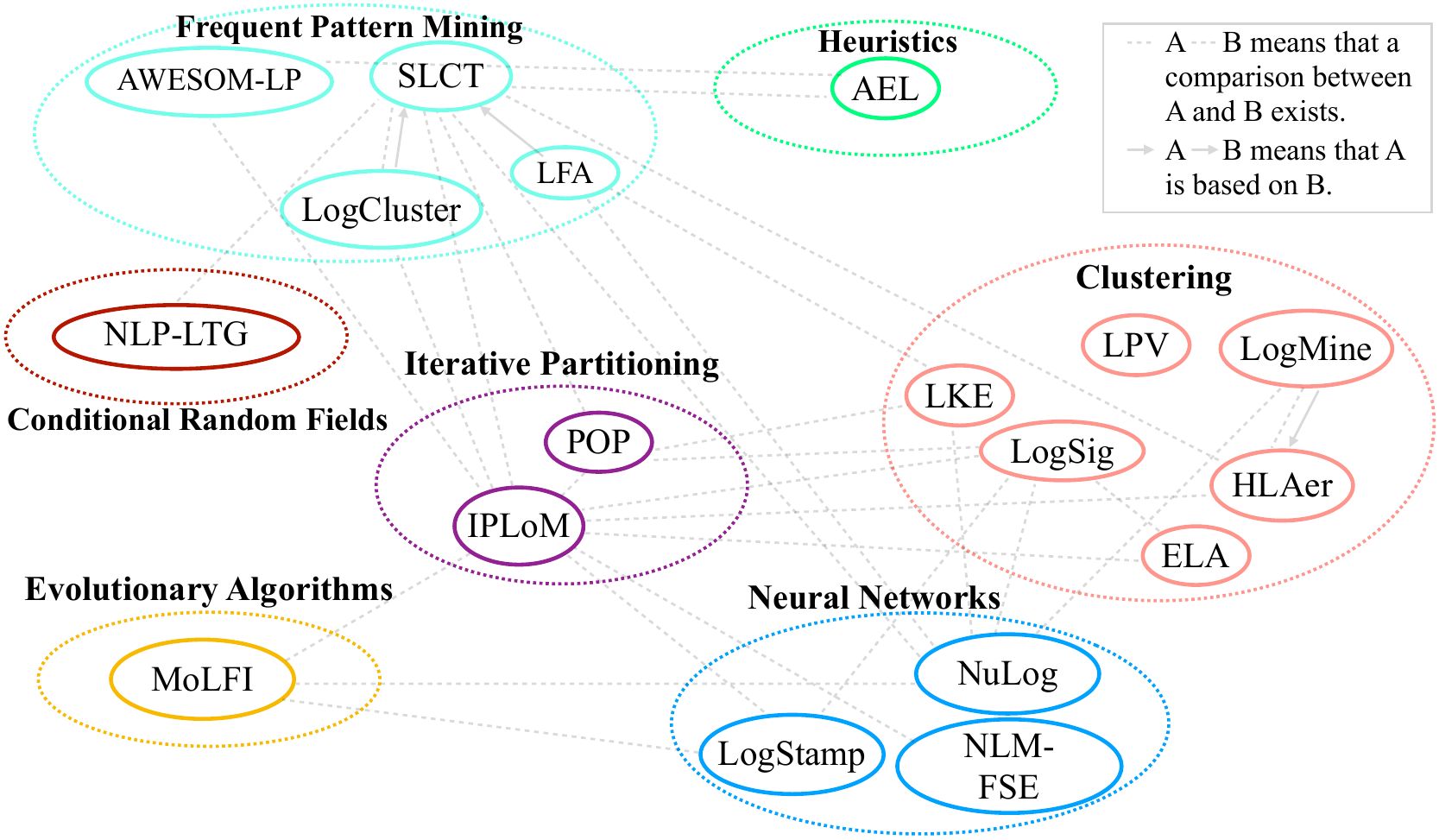}}
 \hspace{0.2in}
	\subfigure{\includegraphics[width=0.36\linewidth]{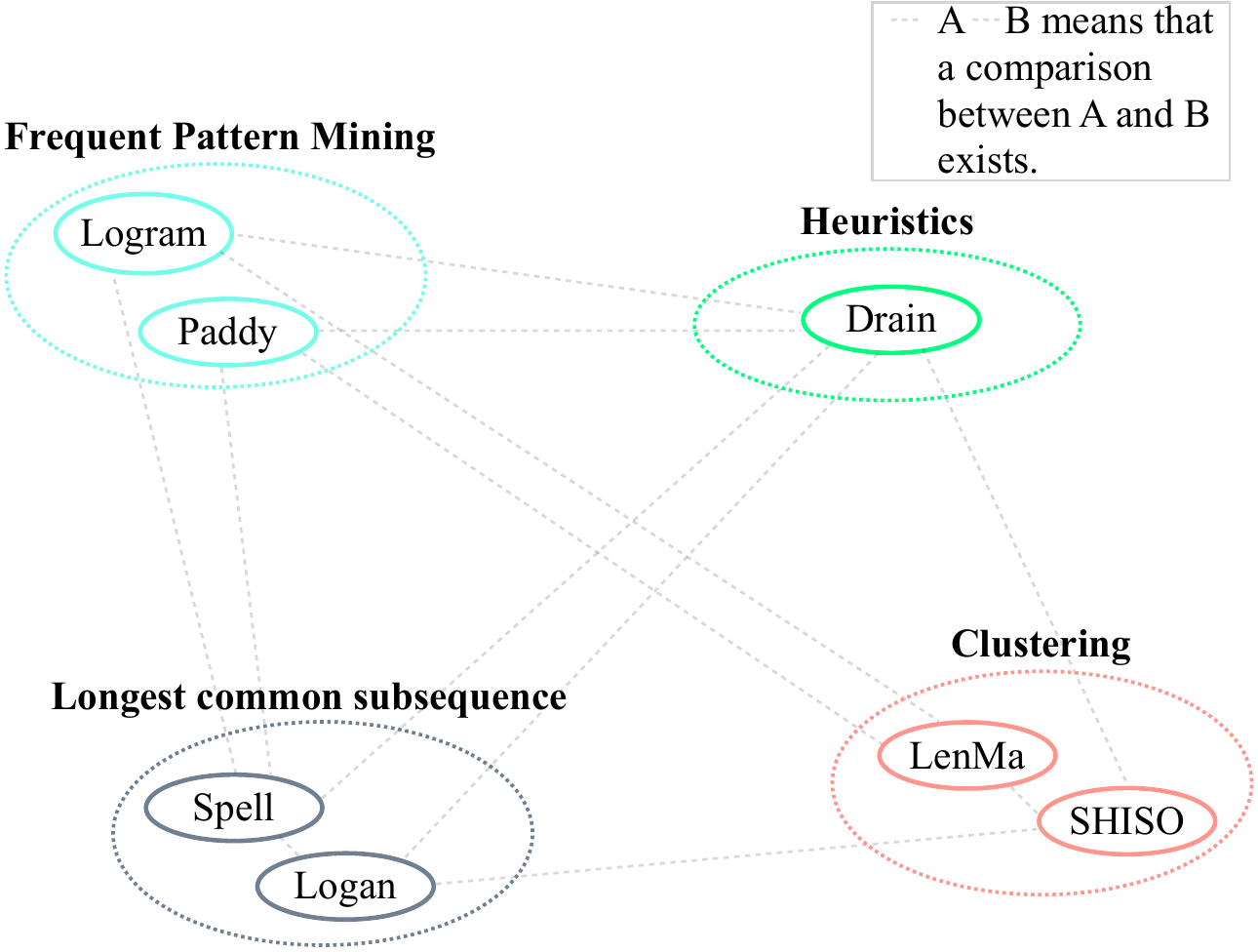}}
	\caption{(a) Clustering of offline approaches. \hspace{0.4in} (b) Clustering of online approaches.}
	\label{fig:offline_online_approaches}
\end{figure*}

Many efforts have gone into log parsing (1), resulting in different algorithms such as IPLoM~\cite{5936060}, LogCluster~\cite{7367331}, LenMa~\cite{shima2016length}, NLM-FSE~\cite{7916497}, Drain~\cite{8029742}, NuLog~\cite{DBLP:journals/corr/abs-2003-07905}, ELA~\cite{10.1145/3373017.3373018}, etc., and (2) surveying log parsing to provide useful overviews of the field~\cite{8804456,10.1145/3460345,DBLP:journals/corr/abs-2110-12489,ELMASRI2020106276,zhangsystem}. 

The different proposed log parsing methods can be grouped into two main categories by the mode in which they process logs ~\cite{ELMASRI2020106276,10.1145/3460345}, namely \emph{offline} and \emph{online}.

\emph{Offline} methods process log data in batches and discover templates given a static set of log messages. They require a training phase, during which the templates are discovered. Subsequently, they parse incoming logs by matching with the templates found during training in either batch or stream~\cite{ELMASRI2020106276}. As changes/updates in software can use new log templates, one drawback of offline approaches is that it requires the training phase to be re-run periodically.

\emph{Online} methods process log data item by item in a streaming manner, and do not require a batch of data to be available before executing. These (methods) discover log templates without an offline training phase, and as log templates are being updated dynamically, such methods can be integrated seamlessly for downstream tasks~\cite{10.1145/3460345}. Online parsers are recommended when the decision time is relatively short (e.g., trying to predict incidents in a software system) and logs need to be processed on the fly.

In Figure~\ref{fig:offline_online_approaches}, we display the log parsing methods proposed in the literature, illustrating how these cluster together based on their underlying algorithmic approach. As we observe in the figure, many methods are designed for parsing logs offline, whereas fewer methods parse logs online. However, from our experience in industry and similarly to what Mahgoub et al.~\cite{273835} claim, we observe that there can indeed be hesitance in industry for applying offline methods, which, although valuable, are mostly avoided by practitioners as they require a significantly higher degree of maintenance, such as constant retraining, and are often times insufficient for real-time use-cases.

\section{A Critique of Log Parsing Evaluation}
\noindent
While log parsing has been studied since the advent of commercial software systems, we claim that the de-facto standard for evaluating log parsing methods provides incomplete performance estimates, bears a strong potential for generating confusion around the role of log parsing, and ultimately creates a disconnect between research and practice. The relevance of this observation is also confirmed by recent studies~\cite{Liu_2022,DBLP:journals/corr/abs-2003-07905}, and we further aim to highlight it: we first reiterate the goals of log parsing, and subsequently discuss metrics that can complement the current de-facto metric toward strengthening the evaluation methodology of the field.

\begin{figure}[t!]
	\centering
	\includegraphics[width=0.9\linewidth]{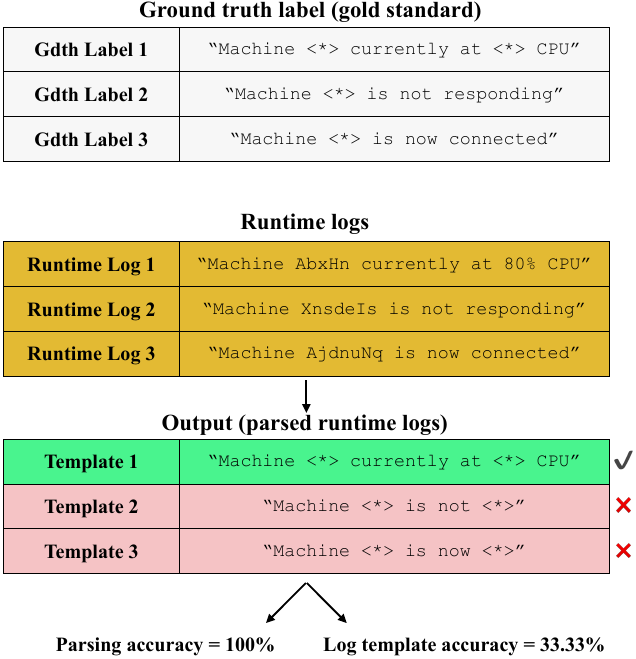}
	\caption{Difference between \emph{parsing accuracy} (considered by He et al.~\cite{10.1145/3460345}) and \emph{log template accuracy}. Even though \texttt{Template 2} and \texttt{Template 3} do not match \texttt{Gdth Label 2} and \texttt{Gdth Label 3} respectively, under the most prevalent metric in the field (\emph{parsing accuracy}), the templates are considered to be extracted correctly, with an accuracy of 100\%.}
	\label{fig:log_template_accuracy_new}
\end{figure}

\subsection{The Choice of Metrics}
\label{sssec:mostprevalentmetric}
\noindent
Log parsing is formulated clearly as the task of mining the underlying log templates that generate runtime logs~\cite{ELMASRI2020106276,9197818,landauer2022deep,zhang2022deeptralog,8067504}. However, a close inspection of the de-facto standard for evaluation --\emph{parsing accuracy}~\cite{8804456}-- reveals that, even though a valid metric, it provides incomplete performance estimates, as this metric disregards the quality of output (parsed logs) completely. Specifically, it assesses the parsers' ability to classify logs, rather than assessing the quality of the log templates that they generate. We illustrate this in Figure~\ref{fig:log_template_accuracy_new}: even though two out of three parsed logs do not match their corresponding ground truth templates, a \emph{parsing accuracy} of 100\% can still be obtained. This is problematic in practice, especially in scenarios where the expected parsed output is to be used, e.g., if variable names are to be extracted and used by downstream processes. We attribute the reason why this disconnect has not been sufficiently researched and addressed so far to the low adoption of log parsing techniques in practice and the marginal degree of automation in actual log analysis pipelines. Overcoming this disconnect between state-of-the-art evaluation methods and actual requirements for log automation, however, is paramount to addressing the needs of industry applications and ensuring that crucial information does not remain undetected due to the manual effort of analysis.

To remove the ambiguity around the goal of log parsing, we (re)define it as the task of identifying log templates and log variables in runtime log messages, and, to ensure that log parsing evaluation also reflects this goal, we consider two evaluation metrics, namely \emph{log template accuracy} and \emph{edit-distance}. We refer to \emph{log template accuracy} as the ratio of the number of correctly parsed logs, over the total number of logs; a log message is parsed correctly if its textual content matches the ground truth template (generated by human experts or mined from software code). By doing so, we elevate the task of log parsing to the same standards as they exist for related problems in information retrieval. 
Secondly, to establish a systematic understanding of the quality of parsing, we rely on the \emph{edit-distance} in the form of the Levenshtein distance~\cite{levenshtein1966binary}. This metric is a more fine-grained alternative to \emph{log template accuracy} as it determines how close the parsed template is from its respective ground truth label. The relevance of these metrics is also confirmed in recent works, where 
\emph{log template accuracy} and \emph{edit-distance} are adopted by Liu et al.~\cite{Liu_2022} and Nedolski et al.~\cite{DBLP:journals/corr/abs-2003-07905}, respectively.

\subsection{Log Parsing Accuracy Scrutinized}
\label{sec:accuracy}
\noindent
Although a large and growing body of literature has investigated the log parsing problem, the implementations of many log parsing methods are not publicly available. Fortunately, Zhu et al.~\cite{8804456} implemented 13 of the most representative approaches in the field, which represents a solid foundation for comparison. Apart from including these 13 in our experiments, we selected one other publicly available method (NuLog~\cite{DBLP:journals/corr/abs-2003-07905}), which resulted in evaluating a total number of 14 log parsing approaches. We display these in Table~\ref{tab:methods_considered_for_experiments}.

\begin{table}[hbt!]
	\centering
	\caption{Log parsing approaches used in our experiments.}
	\label{tab:methods_considered_for_experiments}
	\begin{tabular}{cc|cc}
		\hline
		Year& Method & Year & Method\\
		\hline
		2003 & SLCT~\cite{1251233} & 2015 & LogCluster~\cite{7367331}\\
		2008 & AEL~\cite{4601543} & 2016 & LogMine~\cite{10.1145/2983323.2983358}\\
		2009 & LKE~\cite{5360240} & 2016 & LenMa~\cite{shima2016length}\\
		2010 & LFA~\cite{5463281} & 2017 & Drain~\cite{8029742}\\
		2011 & LogSig~\cite{10.1145/2063576.2063690} & 2018 & MoLFI~\cite{10.1145/3196321.3196340}\\
		2012 & IPLoM~\cite{5936060} & 2019 & Spell~\cite{8489912}\\
		2013 & SHISO~\cite{6649746} & 2020 & NuLog~\cite{DBLP:journals/corr/abs-2003-07905}\\
		\hline
	\end{tabular}
\end{table}

Based on the goals of log parsing and the performance metrics considered in the previous section, parsers are tested on log data in the context of modern software ecosystems. \newline

We consider nine publicly available datasets that have been used extensively in the field for evaluation. Compared to log data generated by modern systems, these contain logs generated within less complex software environments. Consequently, the log parsing methods are expected to perform well. However, our findings indicate that even for homogeneous data (of a single, isolated application), parsers are not able to generate templates that match their respective ground truth templates, and compared to the previous estimates in the field, we discover that the actual performance differs by a large margin. The results of this experiment are summarized in Tables~\ref{tab:accuracy_results_table} and~\ref{tab:edit_distance_results_table}.

\begin{table*}
	\caption{Log template accuracy results after running each method 10 times, for each dataset (2K logs). The measurements are averaged over 10 runs and all standard deviations were below 1\%. We depict in bold font the best results in each category.}
	\centering
	\small
	\setlength\tabcolsep{1.5pt}  
	\label{tab:accuracy_results_table}
	\begin{tabular}{c|ccccccccccccccc}
		\hline
		\textbf{Dataset} & \textbf{AEL} & \textbf{Drain} & \textbf{IPLoM} & \textbf{LenMa} & \textbf{LFA} & \textbf{LKE} & \textbf{LogCluster} & \textbf{LogMine} & \textbf{LogSig} & \textbf{MoLFI} & \textbf{NuLog} & \textbf{SHISO} & \textbf{SLCT} & \textbf{Spell}  \\
		\hline
		Apache & \textbf{0.694} & \textbf{0.694} & \textbf{0.694} & 0.000 & 0.688 & 0.000 & 0.000 & \textbf{0.694} & 0.000 & 0.270 & 0.560 & 0.000 & 0.424 & \textbf{0.694}\\
		BGL & 0.341 & 0.341 & 0.292 & 0.082 & 0.230 & 0.057 & 0.067 & 0.220 & 0.081 & 0.324 & \textbf{0.853} & 0.064 & 0.207 & 0.196\\ %
		HDFS & 0.000 & 0.000 & 0.000 & 0.000 & 0.000 & 0.000 & 0.000 & 0.000 & 0.000 & 0.000 & \textbf{0.435} & 0.000 & 0.000 & 0.000\\
		HealthApp & 0.164 & 0.238 & 0.158 & 0.136 & 0.149 & 0.133 & 0.138 & 0.220 & 0.126 & 0.166 & \textbf{0.341} & 0.041 & 0.322 & 0.152\\
		HPC & 0.644 & 0.620 & 0.638 & 0.632 & 0.609 & 0.360 & 0.632 & 0.632 & 0.509 & 0.632 & \textbf{0.827} & 0.226 & 0.661 & 0.530\\
		Mac & 0.172 & 0.224 & 0.041 & 0.132 & 0.101 & 0.172 & 0.162 & 0.228 & 0.118 & 0.042 & \textbf{0.274} & 0.163 & 0.148 & 0.032\\
		OpenStack & 0.018 & 0.018 & 0.000 & 0.018 & 0.008 & 0.010 & 0.010 & 0.010 & 0.010 & 0.000 & \textbf{0.359} & 0.018 & 0.119 & 0.000\\
		Spark & 0.194 & 0.194 & 0.192 & 0.004 & 
		0.190 & 0.001 & 0.006 & 0.038 & 0.000 & 0.208 & 0.204 & 0.004 & \textbf{0.543} & 0.192\\
		Windows & 0.154 & 0.159 & 0.001 & 0.154 & 0.142 & 0.148 & 0.153 & 0.156 & 0.150 & 0.006 & \textbf{0.387} & 0.151 & 0.140 & 0.004\\
		\hline
		\textbf{Combined Dataset} & 0.267 & 0.258 & 0.214 & 0.140 & 0.180 & 0.140 & 0.128 & 0.258 & 0.092 & 0.180 & \textbf{0.323} & 0.067 & 0.280 & 0.186\\
		\textbf{Industry Dataset} & 0.054 & \textbf{0.056} & 0.041 & 0.001 & 0.022 & 0.001 & 0.002 & 0.054 & 0.000 & 0.048 & 0.050 & 0.002 & 0.034 & 0.041\\
		\hline
	\end{tabular}
\end{table*}

\begin{table*}
	\caption{Edit-distance results after running each method 10 times, for each dataset (2K logs). The measurements are averaged over 10 runs and all standard deviations were below 1\%. We depict in bold font the best results in each category.}
	\centering
	\small
	\setlength\tabcolsep{1.5pt}  
	\label{tab:edit_distance_results_table}
	\begin{tabular}{c|ccccccccccccccc}
		\hline
		\textbf{Dataset} & \textbf{AEL} & \textbf{Drain} & \textbf{IPLoM} & \textbf{LenMa} & \textbf{LFA} & \textbf{LKE} & \textbf{LogCluster} & \textbf{LogMine} & \textbf{LogSig} & \textbf{MoLFI} & \textbf{NuLog} & \textbf{SHISO} & \textbf{SLCT} & \textbf{Spell}  \\
		\hline
		Apache & 10.426 & 10.426 & 10.442 & 13.760 & 10.576 & 14.872 & 16.274 & 10.426 & 14.456 & 10.179 & \textbf{4.679} & 12.648 & 11.234 & 10.442\\
		BGL & 5.014 & 4.930 & 6.882 & 8.373 & 12.524 & 12.582 & 12.955 & 18.598 & 11.921 & 10.969 & \textbf{2.981} & 8.630 & 9.841 & 7.900\\
		HDFS & 8.820 & 8.820 & 16.208 & 10.762 & 30.819 & 17.940 & 28.340 & 16.524 & 18.989 & 19.843 & \textbf{2.867} & 10.114 & 13.641 & 9.274\\
		HealthApp & 19.093 & 18.502 & 11.882 & 16.540 & 20.277 & 28.422 & 16.844 & 19.598 & 17.088 & 21.859 & \textbf{11.595} & 24.430 & 13.840 & 8.540\\
		HPC & 1.405 & 2.015 & 2.323 & 2.906 & 3.182 & 7.649 & 3.580 & 3.218 & 4.419 & 4.845 & \textbf{1.275} & 7.854 & 2.625 & 5.129\\
		Mac & 19.534 & 19.882 & 20.928 & 19.984 & 41.804 & 26.260 & 21.328 & \textbf{17.048} & 28.043 & 28.273 & 21.417 & 19.810 & 34.560 & 22.593\\
		OpenStack & 17.142 & 28.386 & 23.330 & 18.535 & 28.138 & 29.173 & 31.486 & 23.980 & 21.881 & 67.894 & \textbf{5.605} & 18.582 & 20.986 & 27.984\\
		Spark & 3.861 & 3.532 & 5.246 & 10.945 & 9.178 & 18.116 & 17.082 & 16.004 & 12.968 & 14.146 & \textbf{2.921} & 7.910 & 6.028 & 6.129\\
		Windows & 11.975 & 6.172 & 15.758 & 20.662 & 10.238 & 11.834 & 6.967 & 6.919 & 7.667 & 11.943 & 6.067 & 5.624 & 7.006 & \textbf{4.406}\\
		\hline
		\textbf{Combined Dataset} & 13.612 & 17.302 & 14.094 & 18.559 & 24.144 & 27.633 & 21.306 & 15.858 & 24.756 & 19.021 & \textbf{8.721} & 21.791 & 22.274 & 17.454\\
		\textbf{Industry Dataset} & \textbf{21.959} & 27.201 & 24.122 & 32.280 & 41.960 & 68.551 & 47.006 & 23.506 & 37.911 & 49.145 & 23.239 & 31.908 & 46.690 & 24.664\\
		\hline
	\end{tabular}
\end{table*}

\textbf{Finding 1: Previous log parsing evaluations misrepresent the quality of existing solutions.} Previous studies obtained \emph{parsing accuracy} average results of 0.73~\cite{10.1145/3460345}; however, under the considered metric, an average \emph{log template accuracy} of 0.2 is obtained. One might think that an average 0.73 \emph{parsing accuracy} represents matching 1460/2000 templates, whereas, the actual number of templates matched, indicated by \emph{log template accuracy}, is 400/2000. Thus, our results differ from previous estimates substantially and highlight the incompleteness and ambiguity of the de-facto standard for evaluation in the field.

\textbf{Finding 2: Robustness of the method depends on the dataset.} We further discover that, due to the heavily-reliant design on hard-coded rules and heuristics, performance seems to correlate with the particularities of datasets. For example, most methods obtain accuracies of approximately 0.6 for the \emph{Apache} dataset, 0.3 for \emph{BGL}, 0.6 for \emph{HPC}, 0.1 for \emph{Spark}, or 0 for \emph{HDFS}. Nevertheless, we observe that \emph{NuLog}, on the other hand, is significantly more robust with an average accuracy of 0.47. Based on this observation, we argue that it is worthwhile to consider methods that are not heavily reliant on hard-coded rules and heuristics but instead employ methods based on machine learning that have the potential to generalize better in practical applications.

\subsection{The Issue of (Insufficient) Log Complexity}
\label{sssec:insufficient_log_complexity}
\begin{figure*}
	\centering
	\subfigure{\includegraphics[width=0.32\linewidth]{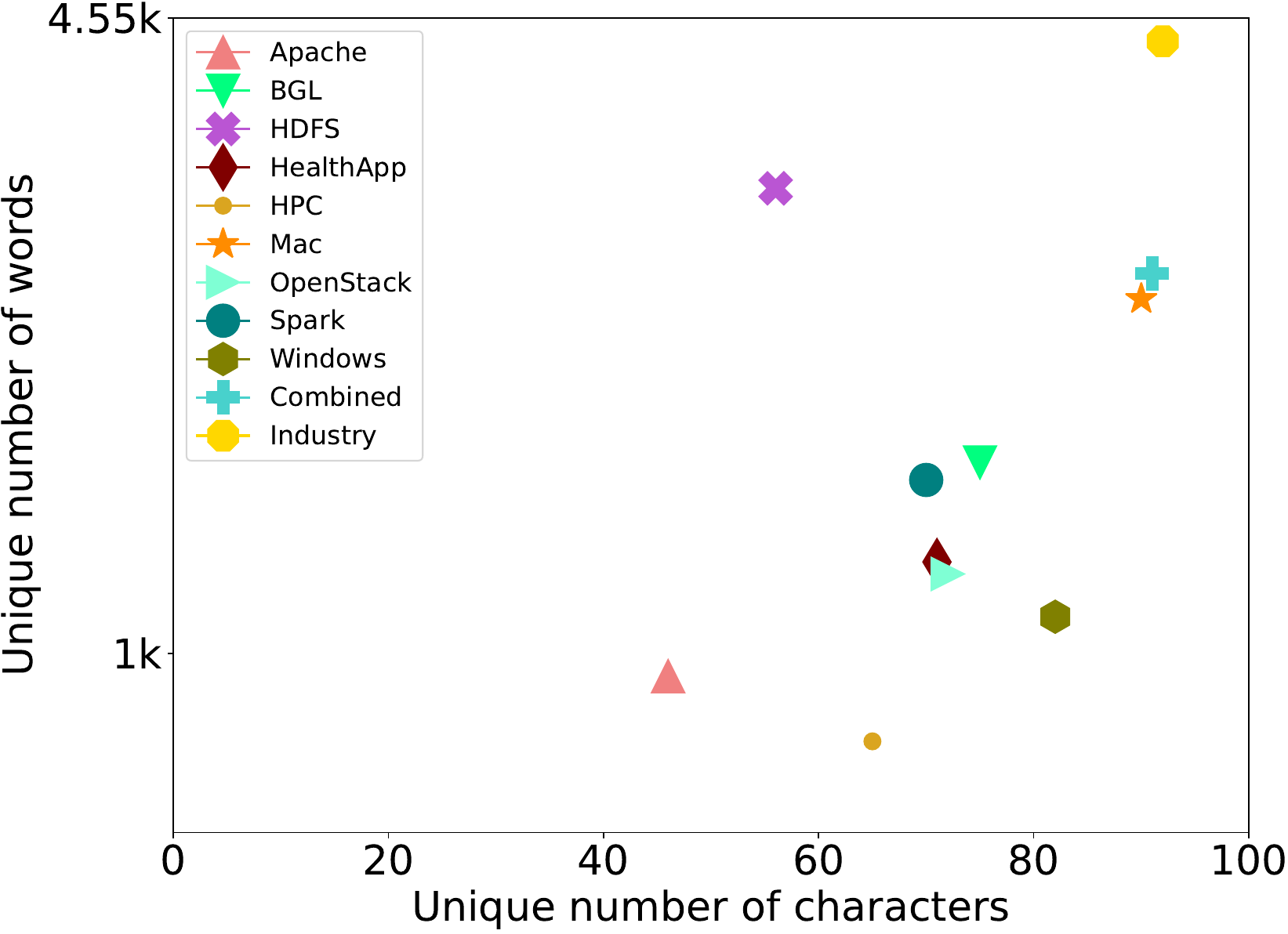}}
	\subfigure{\includegraphics[width=0.32\linewidth]{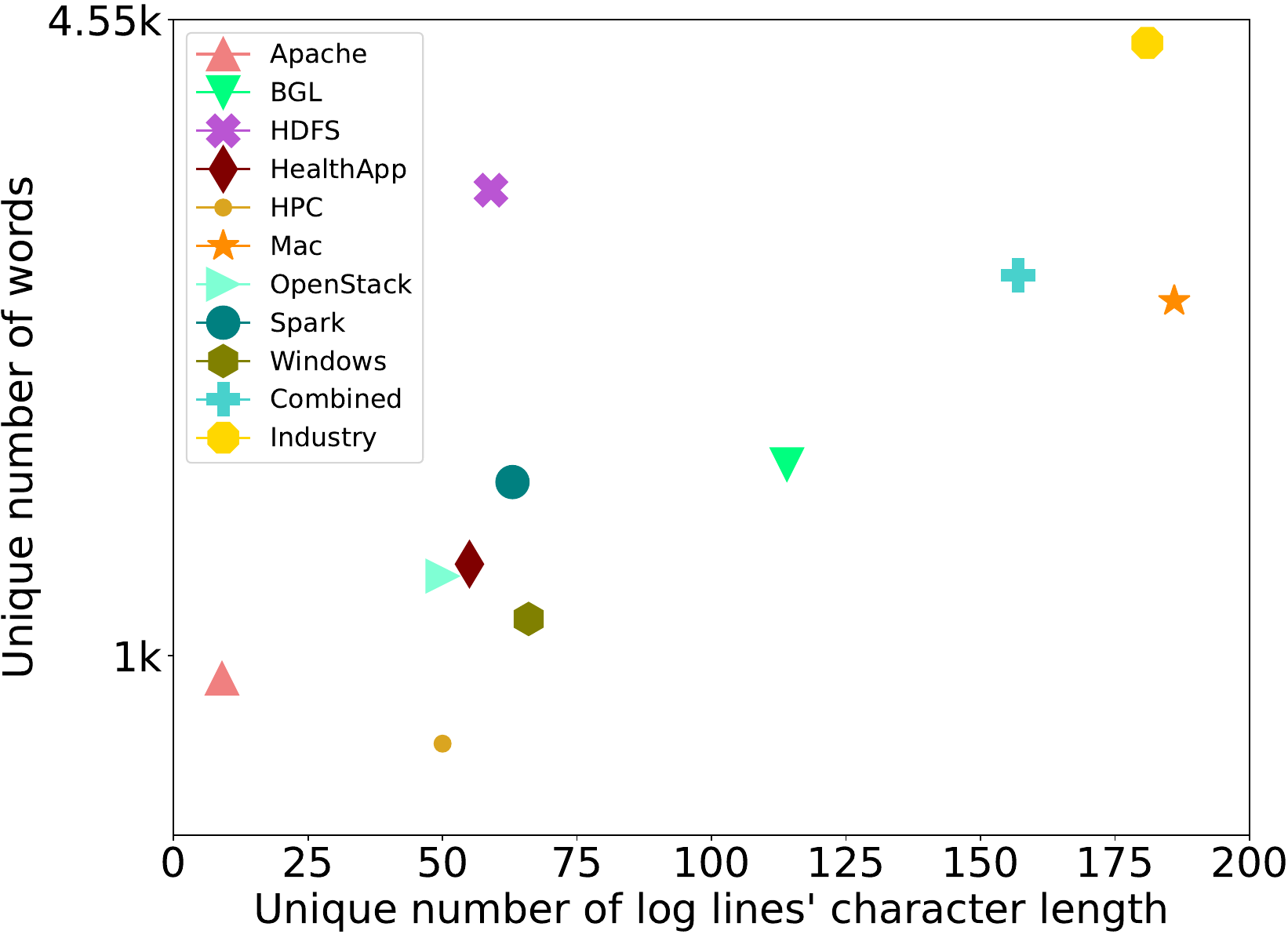}}
	\subfigure{\includegraphics[width=0.31\linewidth]{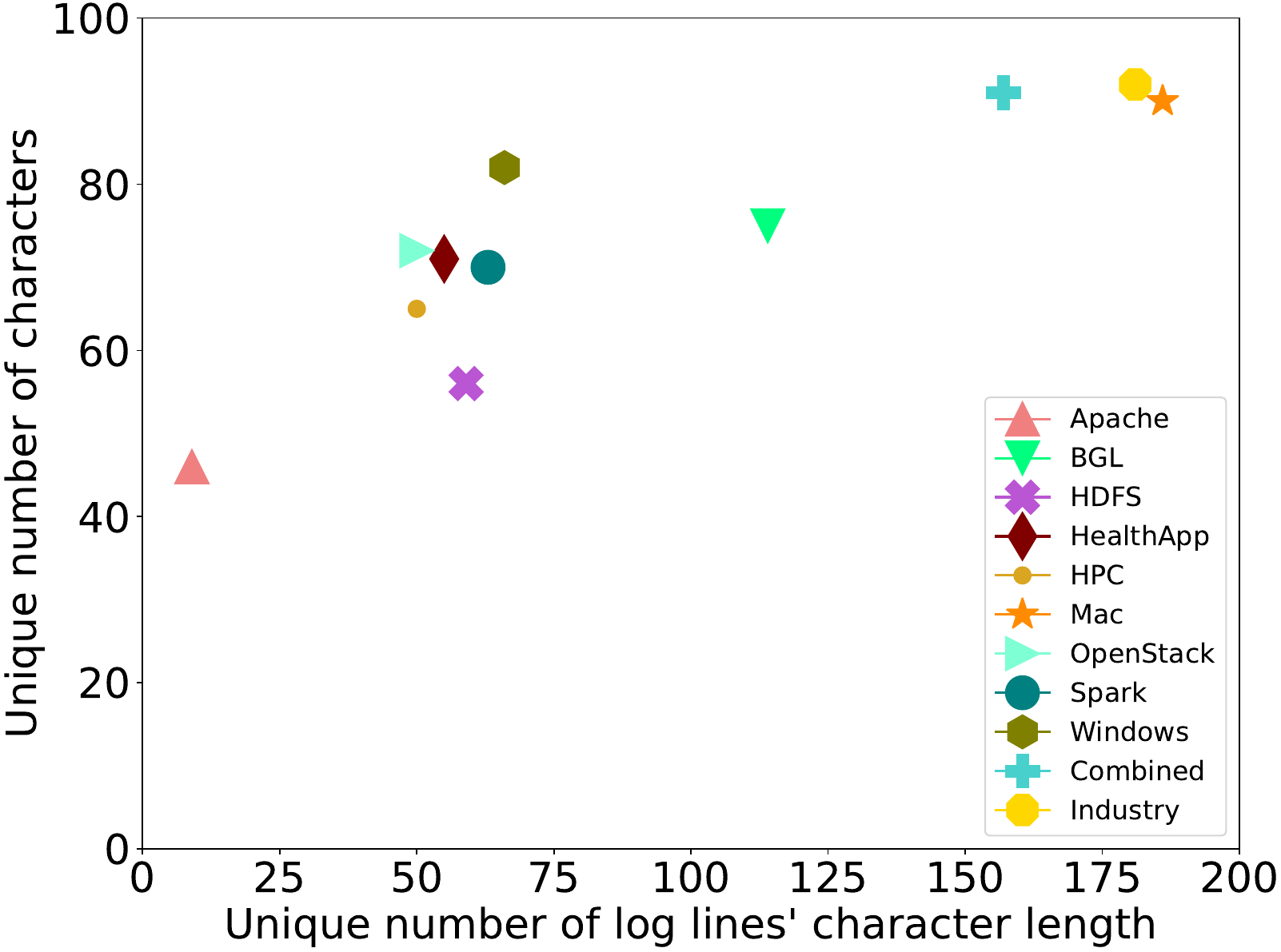}}
	\caption{(a) Unique number of characters versus unique of words. (b) Unique number of log lines' character length versus unique number of words. (c) Unique number of log lines' character length versus unique number of characters.}
	\label{fig:character_versus_words}
\end{figure*}

\noindent
As modern software subsumes various components that generate log data (data engines, processing systems, compute infrastructures, etc.), log heterogeneity is something to expect in industry~\cite{candido2021log,lin2016log}. Consequently, to test parsers on heterogeneous logs, we create a dataset that represents a contemporary software system by drawing a uniform sample across the nine datasets considered in the previous experiment. To demonstrate the similarity between this dataset and real industry data, we consider three metrics that act as a proxy for log heterogeneity; in Table~\ref{tab:statistics_hetero} we present the differences between the log datasets used in our experiments and we visualize these in Figure~\ref{fig:character_versus_words}. We observe that the combined dataset is more reflective of industry compared to the other datasets, as it consistently scores high on all proxy metrics, namely obtaining the highest value for one of the proxy metrics and two second-to-highest values for the other two proxy metrics. Thus, this dataset is considered the baseline for estimating how log parsers perform in practical settings.

\begin{table*}[hbt!]
	\caption{Statistics for the log datasets analyzed.}
	\centering
	\small
	\setlength\tabcolsep{1.5pt}  
	\label{tab:statistics_hetero}
	\begin{tabular}{c|ccccccccccc}
		\hline
		\textbf{Dataset} & \textbf{Apache} & \textbf{BGL} & \textbf{HDFS} & \textbf{HealthApp} & \textbf{HPC} & \textbf{Mac} & \textbf{OpenStack} & \textbf{Spark} & \textbf{Windows} & \textbf{Combined} & \textbf{Industry} \\
		\hline
		No. unique words & 874 & 2068 & 3599 & 1512 & 510 & 2981 & 1445 & 1970 & 1206 & 3123 & 4421\\
		No. unique characters & 46 & 75 & 56 & 71 & 65 & 90 & 72 & 70 & 82 & 91 & 92\\
		No. unique log lengths & 9 & 114 & 59 & 55 & 50 & 186 & 50 & 63 & 66 & 157 & 181\\
		\hline
	\end{tabular}
\end{table*}

We re-run the log parsing methods on the combined dataset and we discover that a comparison between the homogeneous and the heterogeneous case indicates that methods show a marked performance drop. We attribute this result to the limited variability of logs that originate from a specific system: as methods have to parse the same amount of log data as in the previous experiment, but with fewer logs from the same distribution (system), it becomes harder for parsers to recognize variables and to generate quality templates, as illustrated, for example, by the performance drop of \emph{LFA} from scores of 0.688 or 0.230 to 0.180 log template accuracy and from 10.576 or 12.524 to 24.144 edit-distance in Tables~\ref{tab:accuracy_results_table} and~\ref{tab:edit_distance_results_table}, respectively.

Specifically, one assumption when parsing logs is that it is expected for log parsers to distinguish between the constant and variable parts of a message by leveraging access to similar messages that are different only in terms of variables. For example, considering the following log messages: \texttt{`Template log 1'} and \texttt{`Template log 2'}, it is expected for methods to leverage the similarity between these two, and eventually discover the underlying template: \texttt{`Template log <*>'}. However, if log data has more diversity and fewer logs that originate from the same system, the performance takes a substantial hit in terms of \emph{log template accuracy} and \emph{edit-distance}, as the initial assumption is violated and methods are not able to discover patterns in the data as easily.

\textbf{Finding 3: Currently used log datasets are of insufficient complexity to generalize to real-world applications.} The field has not adapted to the rapid changes witnessed over the past decade in the software landscape, which in turn might make current log parsing methods harder to apply, given the change of context. Our results indicate that due to the increase in size and complexity of systems and due to the plethora of subsumed infrastructure resources, the logs generated by these systems have become increasingly more heterogeneous~\cite{candido2021log,lin2016log}, and we showcase that some of the underlying assumptions for applying log parsing may have changed and not hold anymore.

To further exacerbate the problem, current and emerging trends towards more fine-grained componentization (containers, microservices, etc.) will cause these systems to become even more complex in the future~\cite{8354433}, and consequently generate log data of significantly larger volume and increased complexity. Production systems can already create 30-50 Gigabytes of logs per hour~\cite{6410318}, some even reaching terabytes of log data daily~\cite{https://doi.org/10.48550/arxiv.1910.00409}. From our own first-hand experience with industry logs, we observe that contemporary production logs have a significantly higher length compared to logs generated within older monolithic architectures that are commonly used for benchmarks in academic work, which is a consequence of cascaded information from the plethora of components subsumed by the system.

\subsection{Industry Data: Dealing with Real-World Complexity}
\noindent
To evaluate the performance of log parsing techniques in real-world settings, we obtained access to production data through an ongoing collaboration with a major financial institution. Unfortunately, it is rare for academics to have access to such data as it is usually considered sensitive in nature. To make the data usable for evaluation, we had to manually label it with the help of domain experts from the institution who work in job roles that commonly require the analysis of logs for their operational duties. 

As shown in the last rows of Tables~\ref{tab:accuracy_results_table} and~\ref{tab:edit_distance_results_table}, the results show a significant further performance drop from our combined dataset. The highest \emph{log template accuracy} is roughly 0.05, which indicates that the best performing methods parse only 100/2000 log messages correctly, highlighting the difficulty of applying log parsing in production settings. For instance, methods such as LogMine, which previously obtained accuracies as high as 0.694 for publicly available data and 0.258 for combined publicly available data reach an accuracy of only 0.050 on the industry data.

\textbf{Finding 4: Data heterogeneity/diversity is a significant issue for current log parsing methods.} As a consequence of the results obtained on the combined and industry datasets, we argue that data heterogeneity is a significant issue for applying current log parsing methods. To understand the reasons behind the performance drop, we look at that the similarity between the industry dataset and the combined dataset and observe that it is higher than the similarity between the industry dataset and the individual homogeneous datasets. Specifically, the properties of the data found in the industry dataset are very similar to the properties of the combined dataset, as logs originate (similarly) from different data distributions (systems), as a consequence of being centralized. In comparison to the combined dataset, the log diversity found in the industry dataset is higher, but the properties of the dataset are intrinsically the same (clusters of log data generated by different systems). Compared to the nine homogeneous datasets, the log diversity found in the industry dataset is incomparably higher, as it is generated by a significantly larger number of software components. Consequently, this makes it extremely difficult for parsers to discover the underlying templates on industry data, which is reflected in the \emph{log template accuracy} results, and thus the problem is arguably harder than expected from the results obtained on the combined dataset. In terms of \emph{edit-distance} we observe a drop in performance, which can also be attributed to aspects such as jargon \cite{10.1145/3106237.3117776} and high information denseness. Specifically, compared to publicly available data, for a production log, templates and parameters can be harder to separate and identify. For example, it might be that an error occurs on a specific infrastructure resource, which is then propagated to other resources which concatenate and display similar information. In this case, parameters can contain various alphanumeric characters, cascaded messages/nested templates, and also symbols that make it hard for parsers to generate templates that match the corresponding ground truth labels.

\textbf{Finding 5: Lack of access to real-world data hinders log parsing's robustness to real-world scenarios.} We observe that industry data possess qualities that are not as present in publicly available data, potentially making log parsing prone to failure in production environments. Given the performance of log parsing on heterogeneous data, it might be unfeasible to train and design methods on publicly available data as they do not contain the characteristics of industry logs, and it might lead log parsing to ultimately fail in real-world production scenarios. Thus, to increase robustness in real-world scenarios and enable designing methods given industry data characteristics, data that possesses such characteristics is required.

We acknowledge, however, that having access to real-world industry data will always be problematic, as it is well known that companies are reluctant to share production logs due to various concerns, such as privacy, security, etc. Nevertheless, lack of access to real-world data might threaten the research field's pace of progress and practical relevance, as it corners log parsing and puts it at a disadvantage, i.e., publicly available data is easy/easier to parse, and solving log parsing on these data does not translate well to applying it in real-world production scenarios. To close this gap, we aim to address this issue by creating a tool able to generate data that resembles industry logs, without accessing real-world proprietary data.

\section{\textsc{Logchimera}: A Tool for Evaluation on Heterogeneous Data}
\noindent
To address the lack of data for designing and testing log parsing in real-world production scenarios, we designed and implemented \textsc{Logchimera}, a software tool that acts as a proxy to estimate performance on data that resembles industry logs. Specifically, the tool enables (1) estimating log heterogeneity compared to logs as typically found in real-world industry scenarios, (2) increasing log heterogeneity for a given dataset (via mixing data and fuzzing), and (3) enabling a simple interface for both researchers and practitioners to compare state-of-the-art methods using custom data. The name of the tool is inspired by the mythological creature \emph{chimera},  which symbolizes a fusion or combination of different elements; and in this case, it reflects heterogeneity by enabling bringing together diverse formats from various logs to resemble industry-like contexts. In the following sections, we present the tool by discussing its functionality and evaluating its capabilities.

\subsection{Functional Requirements for \textsc{Logchimera}}
\noindent
As we have observed in the previous sections, it becomes tremendously difficult to parse increasingly heterogeneous data (Findings 1 to 3). Consequently, to enable designing and testing using such data, it is necessary to first have a way to estimate log data heterogeneity for a given dataset (Finding 4). Subsequently, based on the estimated heterogeneity, it is essential to have a method to tackle increasing it further to match complex production logs, as we have observed that publicly available data rarely possess the complexity of industry logs directly (Finding 5). 
However, heterogeneity cannot be created artificially from thin air because it needs to be ensured that the resulting log is still syntactically meaningful and resembles log output from realistic applications. With \textsc{Logchimera}, we present a combination of two methods, mixing and fuzzing, which jointly can approximate the heterogeneity of production logs using publicly available log datasets like the ones we used in the previous evaluation. 

To further facilitate the adoption of log parsing methods in practice, we provide an easy-to-use toolkit where people can simply plug in new datasets or new methods and evaluate the resulting performance under realistic conditions. By producing statistical properties of datasets, we enable industry partners to communicate fundamental metrics about the heterogeneity of their log data and consequently researchers to synthetically generate datasets that resemble these statistical properties. Thereby, we hope to help close the gap between the evaluation of log parsers under lab conditions and their performance in production settings.

Practically, given all of the aforementioned aspects, we consider the following functional requirements for \textsc{Logchimera}:

\noindent \textbf{FR1:} \textbf{Estimate log heterogeneity for a given dataset}: Given a log dataset as input, estimate its heterogeneity toward obtaining an estimate of resemblance to industry real world-production data.

\noindent \textbf{FR2:} \textbf{Create logs that posses industry-like characteristics from publicly available data}: Given a log dataset as input, modify its contents toward increasing its heterogeneity and bringing it closer to industry logs.

\noindent \textbf{FR3:} \textbf{Enable access to a simple interface to test state-of-the-art methods against industry-relevant metrics and data}: Given a log dataset as input and a set of methods to be tested, provide performance estimates for the quality of templates generated and the parsers' capacity to extract variables.

In developing the tool, we believe that factors such as usability, transparency and extendability are crucial for future efforts to compound and for strengthening the connection between log parsing in academia and industry.

\subsection{Estimating log heterogeneity}
\noindent
\textsc{Logchimera} estimates heterogeneity as a function of three proxy metrics, namely (1) \texttt{number of unique words}, (2) \texttt{number of unique characters}, (3) \texttt{number of unique lines of different length}. These proxy metrics are used as they provide a simple while reliable way for understanding data heterogeneity: we consider these metrics as logs resemble natural language, and, intuitively, these metrics offer an estimate for how diverse a set of logs is, for instance, a text that contains a higher number of unique of words will be more diverse.

As it can be inconvenient to account for all three dimensions at once when returning an estimate for the heterogeneity of a particular dataset, to obtain a simple and easy to use estimate of heterogeneity, we add the three dimensions (proxy metrics) together into a weighted sum, resulting in a single metric: \emph{H} (log heterogeneity). This is mainly to provide an intuitive proxy metric for heterogeneity; internally, the tool accounts for all three dimensions separately, and the individual scores can be accessed by the user. 

In computing \emph{H} we assign different weights to each individual proxy metric: weighing them differently is motivated by the empirical observation that some of them are more relevant to how heterogeneous a dataset is than others. For instance, in Figure~\ref{fig:character_versus_words}, although some datasets like \texttt{HDFS}, that are highly homogeneous --essentially containing a single log line only with different variables-- can score high in some categories, in this case \texttt{unique number of words}, while actually not being heterogeneous. Consequently, to test this hypothesis, we compute the variance of the proxy metrics across all 11 datasets in Table~\ref{tab:statistics_hetero}: we argue that if the variance is higher, this means the values are more spread out, resulting in a higher disconnect between publicly available data and industry (as industry scores highest), thus providing a better estimate for heterogeneity. Specifically, data that obtains a lower variance signifies that, for a that particular metric, industry is closer to publicly available data, whereas if the variance is higher, we believe the disconnect is higher. We compute the normalized variance of each individual metric and display the results in Table~\ref{tab:statistics_variance_proxy_metrcs}.
\begin{table}[t!]
  \caption{Variance for proxy metrics that motivate weighing these differently when estimating heterogeneity.}
  \centering
  \small
  \setlength\tabcolsep{1.5pt}  
  \label{tab:statistics_variance_proxy_metrcs}
  \begin{tabular}{c|c|c}
    \hline
    \textbf{No} & \textbf{Metric} & $\sigma^2$\\
    \hline
    1 & No. unique words & 0.278\\
    
    2 & No. unique characters & 0.159\\
    
    3 & No. unique log lengths & 0.331\\
    \hline
  \end{tabular}
\end{table}
To account for the the limited number of datasets used in the analysis, we consider a \texttt{40/20/40} weighing, as we observe that the \texttt{no of unique characters} obtains roughly half of the variance of the other two metrics\footnote{We are aware, however, that this is only an approximation, but finding the ``true'' weighing of the metrics is not the focus of this work.}. Consequently, we consider the following formula for estimating log heterogeneity, $H$:

\[H=0.4*\texttt{nuw}\% + 0.2*\texttt{nuc}\% + 0.4*\texttt{nuldl}\%\]
where $\texttt{nuw}$, $\texttt{nuc}$ and $\texttt{nuldl}$ are \texttt{number of unique words}, \texttt{number of unique characters}, and \texttt{number of unique lines of different length}, respectively.

\subsection{Increasing Log Heterogeneity through Mixing}
\noindent
From analyzing publicly available data, we discover that it is usually the case for these data to contain duplicates, or, in many cases to be generated by only a few amount of templates. For example, the number of ground truth templates for the \texttt{Apache}, \texttt{HDFS}, \texttt{HealthApp}, \texttt{HPC}, \texttt{OpenStack}, \texttt{Spark} and \texttt{Windows} is below 70, which is roughly 3.5\% to the ratio of logs in the dataset, meaning that many logs share the same ground-truth label, resulting in a low score of heterogeneity. Consequently, to increase the heterogeneity in such scenarios, we apply a similar strategy to the one we (manually) used when designing the \emph{Combined Dataset} (Section~\ref{sssec:insufficient_log_complexity}) and automate the process behind it: based on the required heterogeneity for a dataset, we replace a certain amount of logs from the original set and mix in a percentage of heterogeneous data. This, however, does not mean that the resulting dataset is changed entirely, but rather, the previously monolithic application log is augmented to approximate a more diverse and heterogeneous landscape of applications contributing to a shared log, as we typically find it in production settings. 

Practically, for increasing the heterogeneity of a dataset via mixing, we gradually replace a percentage of the most frequent entries (based on computing a frequency map of the templates) until a desired level of heterogeneity is obtained: the entries that are used for replacing the original data are taken from a pool of heterogeneous logs, which was created by running an analysis on the nine available datasets and filtering all logs but the 5\% outliers of each dataset, which resulted in a static pool of approximately 500 log lines over 9 datasets (of 18K log lines in total). This technique is based on the empirical observation that it is usually the case for infrequent data to contribute to increasing the three proxy metrics, e.g., an infrequent log contains words that are not found in other log lines of the dataset. To access mixing a dataset toward increasing its heterogeneity, users simply have to provide (1) the log dataset to be modified, having each entry separated by a newline character and (2) the percentage of data to be replaced in the original dataset, ranging from 0 to 1, which maps to 0\% to 25\% (which results in varying the level of heterogeneity of the mixed dataset).

\begin{table}
  \caption{Performance after increasing heterogeneity via mixing. All datasets contain 2K logs. The measurements are averaged over 10 runs and all standard deviations were below 1\%. The score in parenthesis represents the amount of logs that were replaced through mixing other data, e.g., Apache (5) has had 5\% of its most frequent log lines replaced by heterogeneous logs from other datasets.}
  \centering
  \small
  \setlength\tabcolsep{1.5pt}  
  \label{tab:mixing__experiment_results_table}
  \begin{tabular}{c|c|ccc}
    \hline
    \emph{H} level & \textbf{Dataset} & \textbf{AEL} & \textbf{Drain} & \textbf{IPLoM}\\
    \hline
    0.219 & Apache init & 0.694/10.426 & 0.694/10.426 & 0.694/10.442\\
    0.530 & Apache (5) &  0.653/15.925 & 0.653/15.923 & 0.653/15.369\\
    0.640 & Apache (10) & 0.618/19.596 & 0.618/19.602 & 0.618/19.414\\
    0.737 & Apache (15) & 0.586/23.468 & 0.586/23.495 & 0.586/22.809\\
    0.816 & Apache (20) & 0.552/27.504 & 0.552/27.547 & 0.552/26.877\\
    0.886 & Apache (25) & 0.514/32.265 & 0.514/32.135 & 0.514/31.34\\
    \hline
    0.259 & HPC init & 0.644/1.405 & 0.620/2.015 & 0.638/2.323\\
    0.524 & HPC (5) & 0.605/6.510 & 0.581/7.081 & 0.599/7.257 \\
    0.661 & HPC (10) & 0.566/11.734 & 0.542/12.272 & 0.560/12.301 \\
    0.735 & HPC (15) & 0.525/16.720 & 0.501/17.300 & 0.519/17.124 \\
    0.817 & HPC (20) & 0.486/22.040 & 0.462/22.566 & 0.480/21.809 \\
    0.881 & HPC (25) & 0.447/27.875 & 0.423/28.234 & 0.441/27.349 \\
    \hline
    0.608 & BGL init & 0.341/5.014 & 0.341/4.930 & 0.292/6.882\\
    0.756 & BGL (5) & 0.321/11.004 & 0.321/10.940 & 0.273/12.665\\
    0.833 & BGL (10) & 0.303/15.623 & 0.303/15.495 & 0.254/16.325\\
    0.908 & BGL (15) & 0.285/20.665 & 0.285/20.639 & 0.251/20.970\\
    0.949 & BGL (20) & 0.265/25.665 & 0.265/25.771 & 0.216/25.926\\
    \hline
    0.868 & Mac init & 0.172/19.534 & 0.224/19.882 & 0.041/20.928\\
    0.901 & Mac (5) & 0.169/25.184 & 0.217/25.518 & 0.041/26.410\\
    0.919 & Mac (8) & 0.169/28.507 & 0.217/28.838 & 0.041/29.730\\ %
    \hline
    0.830 & Combined Dataset & 0.267/13.612 & 0.258/17.302 & 0.214/14.094\\
    \hline
    1 & Industry Dataset & 0.054/21.959 & 0.056/27.201 & 0.041/24.122\\
    \hline
  \end{tabular}
\end{table}

\begin{table*}[h]
  \caption{Performance on Apache, HPC, BGL, and Mac after increasing heterogeneity via mixing and fuzzing. Measurements are averaged over 10 runs, all standard deviations were below 1\%. The score in parenthesis represents the amount of logs that were replaced via mixing, e.g., Apache (25) has had 25\% of its most frequent log lines replaced by heterogeneous logs from other datasets.}
  \centering
  \small
  \setlength\tabcolsep{1.5pt}  
  \label{tab:mixing_and_fuzzing_results_tabl}
  \begin{tabular}{c|c|ccccc}
    \hline
    \emph{H} level & \textbf{Dataset} & \textbf{AEL} & \textbf{Drain} & \textbf{IPLoM} & \textbf{LFA} & \textbf{LogMine}\\
    \hline
    0.219 & Apache init & 0.694 & 0.694 & 0.694 & 0.688/10.576 & 0.694/10.426\\
    0.886 & Apache (25) & 0.514/32.265 & 0.514/32.135 & 0.514/31.34 & 0.509/20.243 & 0.515/32.016\\
    1 & Apache (25) fuzzed & 0.078/77.541 & 0.118/76.820 & 0.059/45.555 & 0.231/38.866 & 0.060/82.104\\
    \hline
    0.259 & HPC init & 0.644/1.405 & 0.620/2.015 & 0.638/2.323 & 0.609/3.182 & 0.632/3.218\\
    0.881 & HPC (25) & 0.447/27.875 & 0.423/28.234 & 0.441/27.349 & 0.296/17.434 & 0.436/30.290\\
    0.954 & HPC (25) fuzzed & 0.242/97.492 & 0.260/93.240 & 0.219/45.718 & 0.141/42.085 & 0.236/100.324\\
    \hline
    0.608 & BGL init & 0.341/5.014 & 0.341/4.930 & 0.292/6.882 & 0.230/12.524 & 0.220/18.598\\
    0.949 & BGL (20) & 0.265/25.665 & 0.265/25.771 & 0.216/25.926 & 0.150/22.355 & 0.104/33.728\\
    1 & BGL (20) fuzzed & 0.143/95.117 & 0.143/96.156 & 0.073/62.862 & 0.077/46.389 & 0.016/79.8035\\
    \hline
    0.868 & Mac init & 0.172/19.534 & 0.224/19.882 & 0.041/20.928 & 0.101/41.804 & 0.228/17.048\\
    0.919 & Mac (8) & 0.169/28.508 & 0.217/28.838 & 0.041/29.73 & 0.098/51.167 & 0.224/25.933\\
    1 & Mac (8) fuzzed & 0.018/135.994 & 0.011/116.706 & 0.001/112.063 & 0.013/79.454 & 0.022/207.376\\
    \hline
    0.830 & Combined Dataset & 0.267/13.612 & 0.258/17.302 & 0.214/14.094 & 0.180/24.144 & 0.258/15.858\\
    \hline
    1 & Industry Dataset & 0.054/21.959 & 0.056/27.201 & 0.041/24.122 & 0.022/41.960 & 0.054/23.506\\
    \hline
  \end{tabular}
\end{table*}

The runtime of the mixing process (implemented as a part of a Python3 package), e.g., is in the order of \texttt{300ms} for the Apache dataset used in our previous evaluation when mixing in \texttt{500} log lines (25\% of the size of the original dataset) of heterogeneous log data; the runtime for fuzzing (also implemented as a part of a Python3 package) in this scenario is in the order of \texttt{1.3s}.

For evaluating the capabilities of \textsc{Logchimera} when using mixing, we test log parsing on nine publicly available datasets (the same used for the experiments so far). In displaying the results, we select a subset of four datasets, namely \texttt{Apache}, \texttt{HPC}, \texttt{BGL}, \texttt{Mac}, having an increasing level of heterogeneity (based on the measurements obtained in Table~\ref{tab:statistics_hetero}), ranging from homogeneous to heterogeneous, i.e., from \texttt{Apache init} with \emph{H} 0.219 to \texttt{Mac init} with \emph{H} 0.868; results obtained for the other datasets followed a similar trend, thus we consider the subset representative of all datasets. Subsequently, for this experiment we showcase the results of using mixing on a subset of three log parsing methods, namely \texttt{AEL}, \texttt{Drain} and \texttt{IPLoM}; we chose these specifically, as they cover both online (\texttt{Drain}) and offline (\texttt{AEL}, \texttt{IPLoM}) workflows, are used for evaluation by various other works in the field (see Figure~\ref{fig:offline_online_approaches}) and are amongst the most representative works in the field by number of citations, and as the other methods followed a similar trend.

In Table~\ref{tab:mixing__experiment_results_table} we showcase the effects on performance, in terms of log template accuracy and edit-distance, when mixing in data to publicly available data. We observe that as more and more data is mixed in, the performance deteriorates (confirming Finding 4 again), e.g., HPC (25) decreases its performance compared to its orginal state, HPC initial, from 0.644 to 0.447. As, for each dataset, when sampling logs to be mixed in we discard the entries that originate from the dataset about to be mixed, which consequently results in having less of a pool of sampling points for datasets that are already intrinsically heterogeneous (as they themselves contributed to a greater degree to the original pool of heterogeneous outliers). Lastly, the gradually degrading performance when increasing the number of mixed data showcases further that the measurements get closer and closer to the Combined and Industry datasets respectively. This, however, is to be expected, and fulfills the purpose of mixing: generating data that is more heterogeneous, for parsers to test against.

\subsection{Further Increasing Log Heterogeneity through Fuzzing}
\noindent
Motivated by the fact that after a certain threshold mixing cannot provide any improvements in increasing heterogeneity (e.g., as this might mean replacing the dataset entirely) we consider \emph{fuzzing}, a technique that is known from testing~\cite{10.5555/2354410.2355135} and, in our case, intended to help approximate industry logs even further and increase heterogeneity. As we want to ensure, however, that the structure of the logs remains as unaltered as possible during the transformation (instead of replacing with random data that might not resemble logs) we only replace the variables found in logs. For every log line in a particular dataset considered for fuzzing, its variables are replaced with other variables sampled from a pre-computed pool of data, similar to the pool of data gathered for mixing (we extracted the variables from the nine publicly available datasets considered for the experiments, and create a fixed dataset that acts as a resource for increasing heterogeneity). 

\begin{table}[hbt!]
  \caption{Comparison between \emph{H} level obtained leveraging labeled data versus fully automatically based on parsing results and without requiring access to labels.}
  \centering
  \small
  \setlength\tabcolsep{1.5pt}  
  \label{tab:label_vs_no_label}
  \begin{tabular}{c|c|c}
    \hline
    Dataset  & \emph{H} level gdth & \emph{H} level parsed\\
    \hline
    Apache parsed (25) fuzzed & 1 & 0.960\\
    HPC parsed (25) fuzzed & 0.954 & 0.938\\
    BGL parsed (20) fuzzed & 1 & 1\\
    Mac parsed (8) fuzzed & 1 & 0.906\\
    \hline
  \end{tabular}
\end{table}

The distinction of variables from the static portion of a log line is, by itself, the result of parsing log data. As such, we can fully use the results of the previous step for detecting opportunities for fuzzing and then produce a more complex and heterogeneous version of the log data for further evaluation. Clearly, when doing so, the effectiveness of fuzzing is a function of the performance of the log parser. Therefore, we consider two setups in the experimental evaluation, the automated fuzzing based on previous parsing results and applying fuzzing on a labeled dataset where variables are annotated by an expert. While labeled data might not always be available, it is often a prerequisite for training offline methods and it can serve as a best-case scenario to assess the general potential of fuzzing.

Table~\ref{tab:mixing_and_fuzzing_results_tabl} shows the evaluation of \textsc{Logchimera} when using both mixing and fuzzing. As the results show, fuzzing is capable of closing the gap between public datasets and our complex production logs significantly. Specifically, we observe that after fuzzing, not only does the H level increase significantly, e.g., reaching 1 for \texttt{Apache}, \texttt{BGL} and \texttt{Mac}, but the performance on these datasets resembles the same order of magnitude as the performance on industry, e.g., whereas AEL and Drain obtained roughly 0.25 log template accuracy on the \texttt{Combined Dataset} their performance drops down to 0.018 and 0.011 which is in the same order of magnitude with the performance on Industry, 0.054 and 0.056 respectively. There are, however, cases where even though fuzzing increases heterogeneity significantly, the results do not match the Industry performance, as it can be seen for HPC: although performance deteriorates almost double (from 0.447 to 0.242 for AEL) going from mixed data to fuzzed data, the parsed logs seem to have less complexity, as the industry results are in another order of magnitude, i.e., 0.050. We argue that is likely due to mixing and subsequently fuzzing initial logs that are too homogeneous and thereby doe not contain enough properties like distinctly different variables that make these techniques effective. In these cases, one solution could be to either mix in more data, or consider such scenarios as baselines for the level of difficulty parsers have to face. Alternatively, we observe that when the original data is of higher heterogeneity initially, after mixing and fuzzing, the logs seem to posses the characteristics of industry logs, e.g., we see that for \texttt{BGL} the performance drops to the same order of magnitude for three out of the five methods considered, while for the \texttt{Mac} dataset, the performance drops to the same order of magnitude for all methods, matching production log complexity.

Furthermore, in Table~\ref{tab:label_vs_no_label} we compare the heterogeneity of the datasets in both setups: with or without access to labeled data. Consequently, we use \texttt{Drain} for parsing the four displayed datasets, and use the resulting templates and variables as the basis for mixing and fuzzing. For the former, we leverage the statistics of the leveraged templates to mix in data in places indicating logs are generated by the same template (if generated templates are the same for multiple log lines we prioritize replacing them with mixed heterogeneous data), and observe that we are able to match the heterogeneity of the ground truth labels. E.g., for \texttt{Apache} we obtain parsed \emph{H} level heterogeneity scores that are almost identical; it is expected that the \emph{H} level parsed will always be lower than \emph{H} level ground truth since the log template accuracy is not perfect. Furthermore, we observe the effectiveness of using this process, the statistics generated by the templates are good enough to serve as a basis for enhancing the datasets through mixing and fuzzing. For the latter, however, we observe a greater limitation: the quality of fuzzing is a function of how well the parsers were able to extract variables. This means that we expect that for datasets that are intrinsically heterogeneous initially, it will be harder for this process to show the same level of effectiveness without having better parsers available. For instance, compared to the other datasets, \texttt{Mac} scores slightly lower at 0.906, showcasing the phenomenon. Despite that, we are able to use the parsed variables extracted successfully, and increase the heterogeneity of the mixed dataset, e.g., going from an initial \emph{H} level of 0.219 to 0.960 for \texttt{Apache}.

\section{Threats to Validity}
\noindent (1) For the sake of reproducibility, we relied on the same public datasets criticised for not being representative enough of industry. With  \textsc{Logchimera}, we showed a way how researchers can use such datasets and enhance them in complexity through mixing and fuzzing to approximate the distributions of production logs without needing access to the raw logs. We did not show, however, that our method works equally well with other datasets. %

\noindent (2) While acknowledging that our industry dataset may not fully represent all production logs, we utilized it as a reference point in our study. In our efforts, we chose industry data we considered to be challenging, but also representative. By doing so, we demonstrated the difficulties involved in approximating the complexity of such datasets; and, in some cases, depending on the particularities of the software stack and architecture, alternative industry datasets might be easier or even harder to parse.

\noindent (3) The pragmatic nature of the proxy metrics that underlined the \emph{H} level metric may be a threat, as we were unable to provide conclusive evidence of its accuracy, and as a consequence, without considering the varying contributions of different components in estimating heterogeneity may lead to potentially overemphasizing or underemphasizing certain factors. However, our tool can be extended and we encourage future research efforts to potentially propose and incorporate improved metrics: by allowing for the integration of superior metrics, our approach can be improved to address these limitation.

\section{Discussion \& Conclusion}
\label{lbl_d_c}
\noindent
We have investigated log parsing within the context of modern software systems, in an effort of applying it in the infrastructure of a large financial institute. In our findings, we first discovered that log parsing evaluation misrepresents the quality of existing solutions, and highlighted recently proposed metrics that might strengthen the evaluation methodology, and subsequently showed the performance of parsers under these metrics. This led to discovering that methods' robustness is heavily dependent on the particularities of a given dataset, and that data heterogeneity poses a real problem to parsers. This is especially relevant, as industry logs are typically heterogeneous, thus threatening the applicability of log parsing in practice. Furthermore, we discovered that one of the main reasons for the issue of performing poorly on heterogeneous data is the lack of access to designing and testing on it: lack of access to real-world data hinders log parsing's robustness to real-world scenarios. Consequently, to address the lack of data, we designed and implemented \textsc{Logchimera}, a software tool that acts as a proxy for generating and estimating performance on data that resemble industry logs. We hope that our contributions will create a solid connection between research and industry log parsing, adapting its evaluation to the era of modern software systems.

\newpage

\bibliographystyle{IEEEtran}
\bibliography{sample}

\end{document}